\documentclass[11pt]{article}
\usepackage{euler}
\usepackage{ams}
\usepackage{xypic}
\newtheorem{tm}{Theorem}

\newtheorem{cha}{Challenge}

\input xy
\xyoption{all}

\begin{document}
\begin{center}{\Large{\bf A new Protocol for 1-2 Oblivious Transfer}}\\
\vspace{.5cm}
{Bj\"orn Grohmann}\\\vspace{.2cm}
{\small Universit\"at Karlsruhe, Fakult\"at f\"ur Informatik\\76128 Karlsruhe, Germany}\\
{\small {\tt nn@mhorg.de}}
\end{center}
\vspace{.5cm}
\begin{abstract}
A new protocol for 1-2 (String) Oblivious Transfer is proposed. The protocol uses 5 rounds of message exchange.\\\\
{\bf Keywords:} Oblivious Transfer, cryptographic Hash-Function, One-Way-Function.\\ 
\end{abstract}
\section{Introduction}
During a 1-2 (String) Oblivious Transfer protocol, Bob should learn one of two bit strings provided by Alice, but not both, while Alice should not learn anything about Bob's choice.\\\\
A protocol fulfilling these constraints would be a powerful cryptographic primitive (cf. \cite{killian88} for an introduction to the subject).\\\\
In this article, we propose a protocol that uses 5 rounds of message exchange. Since most of the computational part of the protocol takes place in the unit group of a finite field, we further investigate the question whether Alice or Bob can gain more information, if it turns out that the computation of discrete logarithms in this group is easy.
\section{The Protocol}
{\bf Initialisation:} Before the actual start of the protocol Alice and Bob agree on a positive integer $n\in\N$, a prime $p$ of size $\sim 2^{\sqrt{n\log n}}$, a random matrix $C=\left(c_{i,j}\right)_{i,j}\in\F_p^{n\times n}$, $i,j=1,\dots n$, a cryptographic Hash-Function
$h_1:\F_p\longrightarrow \{0,1\}^q$ and an injective (polynomial-time computable) One-Way-Function 
$h_2:\{0,1\}^q\longrightarrow \{0,1\}^{q^{\prime}}$, for integers $q$ and $q^{\prime}$. Here, $\F_p$ denotes the finite field with $p$ elements.\\\\
{\bf Round 1:} Alice starts by choosing $n$ random bits $t_1,\dots,t_n$, two distinct random elements $a,b\in\F_p$, with $a\not= -c_{i,j}\not= b$ for $i,j=1,\dots,n$, two distinct random elements $\alpha_a,\alpha_b\in\F_p^{\times}$ of order $p-1$ (i.e. each of these elements is a generator of the unit group $\F_p^{\times}:=\F_p-\{0\}$) and two random permutations $\sigma_a,\sigma_b$ on the set $\{1,\dots,n\}$. She then computes, for $j=1,\dots,n$,
\begin{equation}\label{eqmu}
\mu_{j,a} := \alpha_{a}^{\sigma_{a}(j)}\prod_{i=1}^{n}\left(a+c_{i,j}\right)^{t_i}\,\,\mbox{ and }\,\,\,\,
\mu_{j,b} := \alpha_{b}^{\sigma_{b}(j)}\prod_{i=1}^{n}\left(b+c_{i,j}\right)^{t_i}
\end{equation}
and sends $((\mu_{j,a})_j,(\mu_{j,b})_j)$ to Bob.\\\\
{\bf Round 2:} Bob chooses $n$ random bits $s_1,\dots,s_n$. He computes
\begin{equation}
\tau_{A,a} := \prod_{j=1}^{n}\mu_{j,a}^{s_j}\,\,\,\mbox{ and }\,\,\,\tau_{A,b} := \prod_{j=1}^{n}\mu_{j,b}^{s_j}
\end{equation}
and sends $(\tau_{A,a},\tau_{A,b})$ to Alice.\\\\
{\bf Round 3:} Alice chooses two (random) bit strings $m_a$, $m_b$ of size $q$ (the messages) and computes $z_a:=h_2(m_a)$ and $z_b:=h_2(m_b)$. She then computes, for $k=1,\dots,\frac{n(n-1)}{2}$,
\begin{equation}
s_{k,a} := h_1\left(\alpha_a^{-k}\tau_{A,a}\right)\oplus m_a\,\,\,\mbox{ and }\,\,\,s_{k,b} := h_1\left(\alpha_b^{-k}\tau_{A,b}\right)\oplus m_b,
\end{equation}
where $\oplus$ denotes the {\tt XOR}-function, and sends $((s_{k,a})_k,(s_{k,b})_k,a,b,z_a,z_b)$ to Bob.\\\\
{\bf Round 4:} Bob chooses a random element $\beta\in\F_p^{\times}$ of order $p-1$, a random permutaion $\varrho$ on the set $\{1,\dots,n\}$ and an element $d\in\{a,b\}$. He then computes, for $i=1,\dots,n$,
\begin{equation}
\nu_i := \beta^{\varrho(i)}\prod_{j=1}^{n}(d+c_{i,j})^{s_j}
\end{equation}
and sends $(\nu_i)_i$ to Alice.\\\\
{\bf Round 5:} Alice computes
\begin{equation}\label{eq5}
\tau_B := \prod_{i=1}^{n}\nu_i^{t_i}
\end{equation}
and sends $\tau_B$ to Bob.\\\\
Finally, Bob computes for $r=1,\dots,\frac{n(n-1)}{s}$ the list $(\beta^{-r}\tau_B)_r$ until he finds $r_0$ and $k_0$ such that
$h_2(h_1(\beta^{-r_0}\tau_B)\oplus s_{k_0,d})=z_d$,
which gives him the message $m_d =h_1(\beta^{-r_0}\tau_B)\oplus s_{k_0,d}$.
\section{Analysis}
The following theorem states the correctness of the protocol and (roughly) counts the computational cost for both sides (for simplicity, we count addition and multiplication in $\F_p$ as one elementary operation and leave aside the randomized selection process).
\begin{tm}
At the end of the protocol, Bob is in possession of the message he asked for. The computational cost for Alice equals ${\bf O}(n^2\cdot(\mbox{{\rm cost of }}h_1))$ elementary operations, while on Bob's side it sums up to ${\bf O}(n^2\cdot(\mbox{{\rm cost of }}h_1)+n^4\cdot(\mbox{{\rm cost of }}h_2))$.
\end{tm}
{\bf Proof.} The first statement of the theorem is easily seen to be true, since
\begin{equation}
\tau_{A,d} = \alpha_d^{k^{\prime}}\prod_{i,j=1}^n (d+c_{i,j})^{t_i s_j}
\end{equation}
and respectively
\begin{equation}
\tau_B = \beta^{r^{\prime}}\prod_{i,j=1}^n (d+c_{i,j})^{t_i s_j},
\end{equation}
with $d\in\{a,b\}$ and $1\leq k^{\prime},r^{\prime}\leq n(n-1)/2$. The calculation of the computational cost is straightforward.
\hfill$\Box$\\\\
We now turn to the two fundamental questions for this protocol. For this, we define the function $f(y):=\prod_{i,j}(y+c_{i,j})^{t_i s_j}$. It is clear that, for $d\in\{a,b\}$, the knowledge of $f(d)$ leads to the knowledge of the message $m_d$.\\\\
{\bf Q1:} Can Alice efficiently decide whether Bob chose $d=a$?\\\\
{\bf Q2:} Can Bob, who knows $f(d)$, efficiently compute $f(a+b-d)$?\\\\
So far, the author of this article is not aware of any polynomial time algorithm that would answer one of these questions with ``yes''.\\\\
In the following we shall see that even the ability to efficiently compute discrete logarithms in $\F_p^{\times}$ does not seem to help much.\\\\
So, from now on we will assume that Alice and Bob can compute discrete logarithms in $\F_p^{\times}$ efficiently. To start with Bob (i.e. {\bf Q2}) it is easily seen that the knowledge of Alice's secret bits $t_1,\dots,t_n$ immediately gives him both messages $m_a$ and $m_b$ (he can compute $f(a)$ and $f(b)$). To get these bits, Bob can choose a generator $g$ of the group $\F_p^{\times}$ and try to solve the equation (cf. (\ref{eq5}))
\begin{equation}\label{eqdl}
x_1 \delta_g(\nu_1)+\cdots+x_n \delta_g(\nu_n) \equiv \delta_g(\tau_B)\,\,{\rm mod}\,\,p-1,
\end{equation}
where $\delta_g(\cdot)$ denotes the discrete logarithm function with respect to $g$. Since there are $2^n$ ways to select the values of the $x_i$'s, there are, heuristcally speaking, approximately $2^{n-\log p}\sim 2^{n(1-\sqrt{\log n/n})}$ solutions to equation (\ref{eqdl}). Now suppose that Bob knows $f(a)$. He then can compute $\alpha_a^{k^{\prime}}$, with an unknown positive integer $k^{\prime}\leq n(n-1)/2$. Suppose further that he somehow manages to determine $\alpha_a$ (or at least a list of possible candidates for $\alpha_a$). Since ${\rm gcd}(\delta_q(\alpha_a),p-1)=1$ this leads (cf. (\ref{eqmu})) in general to the following
\begin{cha}\label{cha1}
Given $n\in \N$, a prime $p$ of size $\sim 2^{\sqrt{n\log n}}$, a matrix $\left(e_{i,j}\right)_{i,j=1,\dots,n}$ with integer coefficients and a list of integers $(f_j)_{j=1,\dots,n}$, compute $x_1,\dots,x_n$, with $x_i\in\{0,1\}$, and a permutation $\pi$ on the set $\{1,\dots,n\}$ such that
\begin{eqnarray*}
x_1 e_{1,1}  +  \dots  +  x_n e_{1,n}  +  \pi(1) & \equiv & f_1\,\,{\rm mod }\,\,p-1\\
x_1 e_{2,1}  +  \dots  +  x_n e_{2,n}  +  \pi(2) & \equiv & f_2\,\,{\rm mod }\,\,p-1\\
& \vdots & \\
x_1 e_{n,1}  +  \dots  +  x_n e_{n,n}  +  \pi(n) & \equiv & f_n\,\,{\rm mod }\,\,p-1.\\
\end{eqnarray*}
\end{cha}
Again, the author of these lines is not aware of any efficient method that solves this challenge.\\\\
Now, Alice's story ({\bf Q1}) is pretty much the same. In the end, Alice finds herself confronted with a decision version of Challenge \ref{cha1}, but as is easily seen,
an algorithm that can decide in polynomial time whether a solution exists can also be used to efficiently compute a solution.

\end{document}